\documentclass[prd,aps,preprint,preprintnumbers,nofootinbib,showpacs,amsmath,amssymb,superscriptaddress]{revtex4-1}
\usepackage{color}
\usepackage{graphicx}

\usepackage{amsmath, amssymb}
\usepackage{fontenc}
\usepackage[utf8]{inputenc}
\usepackage{lmodern}
\usepackage[english]{babel}

\usepackage[normalem]{ulem}

\newcommand{\simgt}{\lower.5ex\hbox{$\; \buildrel > \over \sim \;$}}
\newcommand{\simlt}{\lower.5ex\hbox{$\; \buildrel < \over \sim \;$}}

\begin{document}

\title{Constraints on $\alpha$-attractor inflation and reheating}
%

%
\author{
Yoshiki Ueno$^{1}$ and Kazuhiro Yamamoto$^{1,2}$
}

\date{\today}
\affiliation{
$^{1}$Department of Physical Science, Graduate School of Science, Hiroshima University, \\ Kagamiyama 1-3-1,
         Higashi-hiroshima 739-8526, Japan \\
$^{2}$Hiroshima Astrophysical Science Center, Hiroshima University,  \\Kagamiyama 1-3-1,
         Higashi-hiroshima 739-8526, Japan }

\begin{abstract}%
We investigate a constraint on reheating followed by $\alpha$-attractor-type 
inflation (the E-model and T-model) from an observation of the spectral index $n_s$. 
When the energy density of the universe is dominated by an energy component with
the cosmic equation-of-state parameter $w_{\rm re}$ during reheating, its $e$-folding 
number $N_{\rm re}$ and the reheating temperature $T_{\rm re}$ are bounded depending on $w_{\rm re}$. 
When the reheating epoch consists of two phases, where the energy density of 
the universe is dominated by uniform inflaton field oscillations in the first phase and 
by relativistic non-thermalised particles in the second
phase, we find a constraint on the $e$-folding number of the first oscillation 
phase, $N_{\rm sc}$, depending the parameters of the inflaton potential.
For the simplest perturbative reheating scenario, we find the lower bound for a
coupling constant of inflaton decay in the E-model and T-model depending on
the model parameters. We also find a constraint on the $\alpha$ parameter, 
$\alpha\simgt 0.01$, for the T-model and E-model when we assume a broad resonance 
reheating scenario.
\end{abstract}

\maketitle
\section{Introduction}
Inflation is a key to exploring the beginning of the universe.
There are various inflation models. However, recent precise observations of 
the cosmic microwave background and the large-scale structure of galaxies
impose useful constraints on inflation models \cite{Planck}. 
The combination of constraints on the spectral index $n_s$ and the scalar tensor
ratio $r$ excludes the simplest single power-law potential models. 
A class of inflation models that is consistent with observations is 
the $\alpha$-attractor-type models, which were recently proposed in a unified manner
\cite{SCI1,SCI2,SCI3,SCI5,SCI6,Galante,Scalisi,Linde2015,Scalisi2}; they include the Starobinsky model
\cite{Starobinsky,Mukhanov} (cf., \cite{Nariai,NariaiTomita,Nariai3,Nariai4}) and the Higgs 
inflation model \cite{Salopek,Bezrukov,Okada,Bezrukov2,LNW}. 

Reheating after inflation is important for the inflation model itself as a mechanism to 
realise the hot big bang universe. The energy of an inflaton field is converted to 
thermal radiation during a reheating epoch by processes that may include the physics of 
particle creation and non-equilibrium phenomena. 
Reheating processes have been investigated in many studies 
(e.g., \cite{Abbott,Dolgov,Albrecht,Trashen,Kofman,Shtanov,KLS,Mukhanovtext}), 
in which successful scenarios of preheating and subsequent thermalisation 
processes were discussed; however, many uncertainties still remain 
(see, e.g., \cite{Allahverdi,Kaiser} for a review). 

Some authors recently investigated a constraint on the reheating epoch 
\cite{Dai,Munoz,Cook} that uses a recent precise constraint
on the spectral index $n_s$ \cite{Planck}. 
The authors of \cite{Dai,Munoz,Cook} investigated constraints on 
the $e$-folding number and reheating temperature depending on 
the effective equation-of-state parameter of the reheating epoch.
In this paper, we investigate the constraint on the reheating epoch 
of the $\alpha$-attractor-type inflation models.
The authors of \cite{Cook} 
investigated the constraint on the reheating epoch in the Higgs inflation 
model ; however, our investigations focus on a wider class of  
$\alpha$-attractor-type models, the E-model and T-model \cite{Carrasco1,Carrasco2}, 
which are consistent with the observations. 
Some aspects of reheating followed by the E-model and T-model was investigated 
in Ref.~\cite{Cai} by introducing a phase diagram, but we
examine this problem from a different perspective. 

In our investigation, our approach to a constraint on 
reheating differs from those of \cite{Dai,Munoz,Cook,Cai}. These previous
works assume that the universe is dominated by an energy density with 
a constant equation-of-state parameter $w_{\rm re}$. 
In this work, we consider a reheating epoch consisting of two phases. 
The first phase is an epoch in which the energy of the universe is dominated 
by uniform inflaton field oscillations (the oscillation phase), and the second 
phase is an epoch in which the universe is dominated by relativistic but 
non-thermalised particles produced by decay of the inflaton field 
(the thermalisation phase). 
Our analysis constrains the $e$-folding number of the oscillation phase 
using an observation of the spectral index $n_s$, which we use to discuss 
constraints on a parameter of the inflaton potential and a coupling constant 
for inflaton decay depending on two reheating scenarios. 

This paper is organised as follows: 
In section 2, we briefly review how to constrain the 
$e$-folding number of reheating and the reheating temperature 
using an observational constraint on the spectral index $n_s$.
In section 3, we investigate a constraint on the reheating epoch 
in a single-field $\alpha$-attractor model, assuming that the reheating
epoch is dominated by an energy component with the equation-of-state 
parameter $w_{\rm re}$. We demonstrate that our result is consistent 
with previous results.
In section 4, we consider a reheating epoch consisting of
the two phases, i.e., the scalar field oscillation phase and the thermalisation 
epoch. In section 5, we discuss the impacts of our results on two 
reheating scenarios. 
Section 6 presents a summary and conclusions. We adopt 
the convention $M_{\rm pl}^2=1/8\pi G$, where $G$ is the 
gravitational constant.

\section{Constraint on reheating}
\begin{figure}[t]
 \begin{center}
    \hspace{0mm}\scalebox{0.5}{\includegraphics{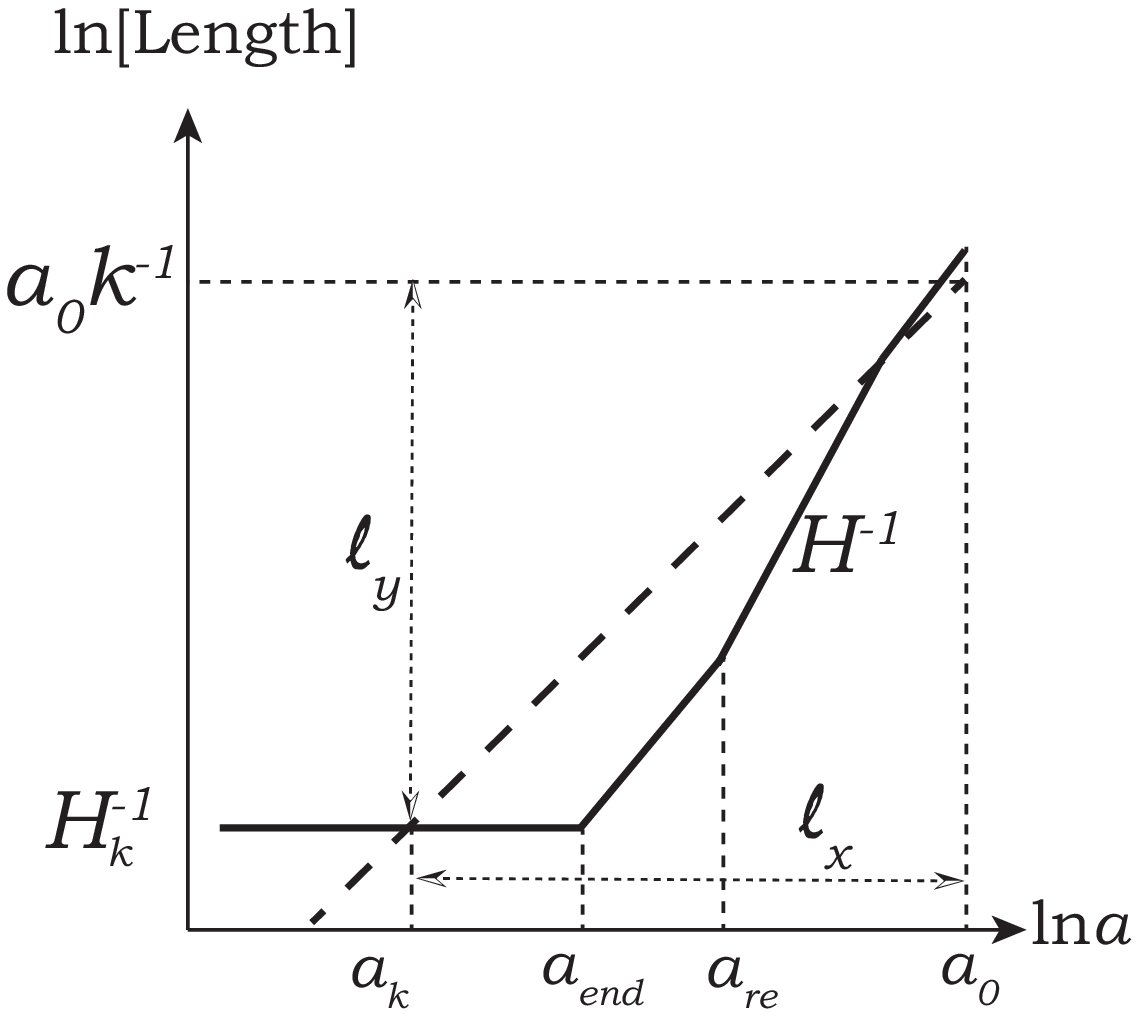}}
    \hspace{1cm}
    \hspace{0mm}\scalebox{0.5}{\includegraphics{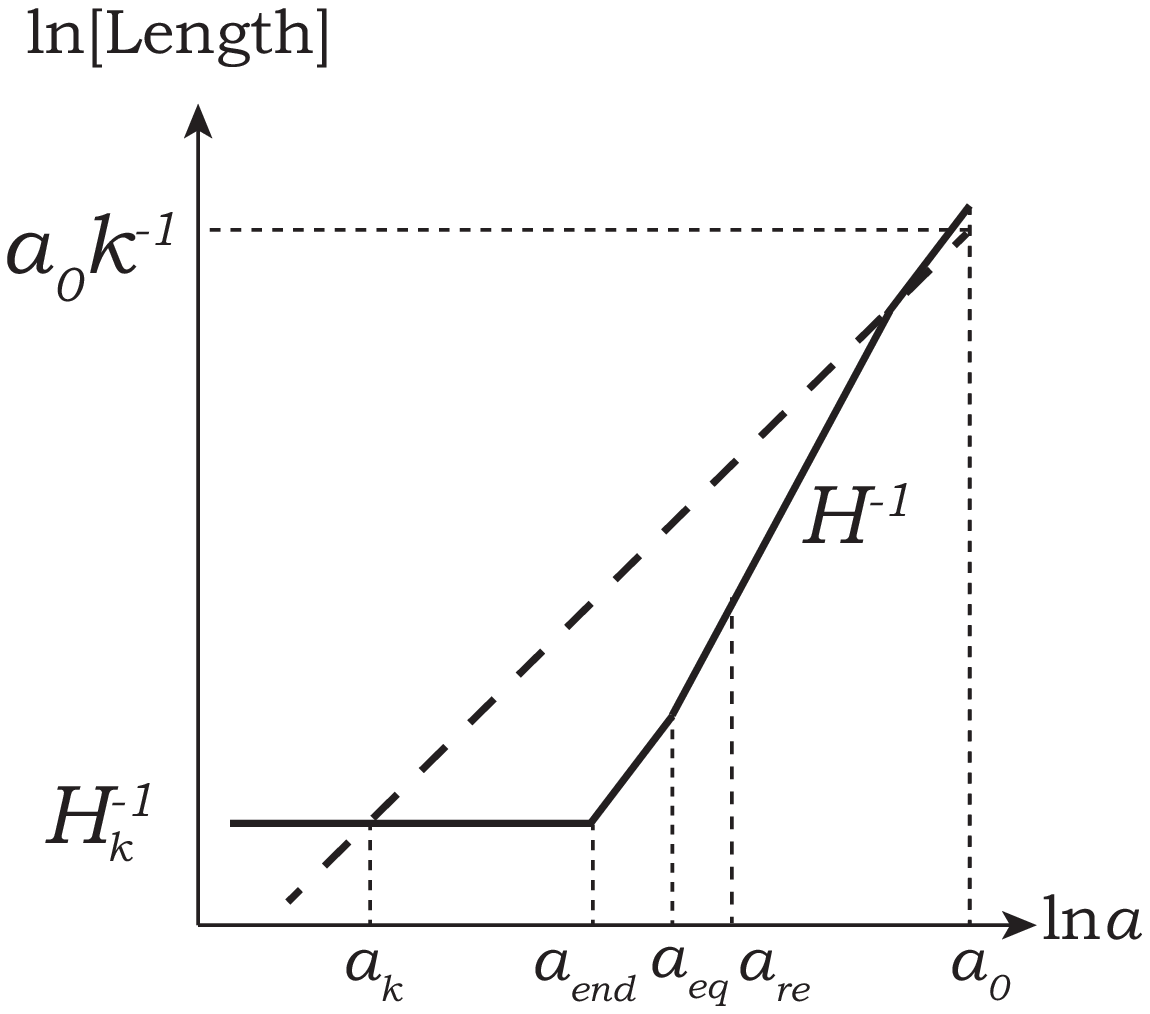}}
 \caption{Sketch of evolution of Hubble horizon distance $H^{-1}$ (solid curve) 
from the inflation epoch to the present epoch as a function of the scale factor $a$. 
Long dashed line shows the evolution of the physical wavelength of a perturbation 
with the comoving wavenumber $k$. 
Here a logarithmic scale is adopted for both axes. 
$a_k$, $a_{\rm end}$, $a_{\rm eq}$, $a_{\rm re}$, and $a_0$ are the scale factors at 
horizon crossing during inflation, at the end of inflation, at the equal time 
during reheating, at the end of reheating, and at the present epoch, respectively.
Left and right panels illustrate the assumptions in sections 3 and 4, respectively. 
In the present paper, we adopt $k=0.05{\rm Mpc }^{-1}$. 
\label{figeras}}
 \end{center}
 \end{figure}

We briefly review how to constrain the $e$-folding number of reheating 
$N_{\rm re}$ and the reheating temperature $T_{\rm re}$ using an observational 
constraint on the spectral index $n_s$ \cite{Dai,Munoz,Cook,Cai}. 
We consider a single-field inflation model with a potential $V(\phi)$, 
which obeys
\begin{eqnarray}
&&\ddot\phi+3{\dot a\over a}\dot\phi+{\partial V\over \partial \phi}=0,
\label{fieldequations2}
\end{eqnarray}
where the dot indicates differentiation with respect to
cosmic time, and $a$ is the scale factor
determined by the Friedman equation:
\begin{eqnarray}
&&\left({\dot a\over a}\right)^2
  ={1\over 3M_{\rm pl}^2}\left({\dot\phi^2\over 2}
  +V(\phi)\right).
\label{fieldequations1}
\end{eqnarray}
Adopting the slow-roll approximation during inflation,
the above equations are approximated as
\begin{eqnarray}
&&3H\dot\phi+V'(\phi)=0,
\\
&&H^2={V(\phi)\over 3M_{\rm pl}^2},
\end{eqnarray}
where the prime denotes differentiation with respect to $\phi$, and 
$H=\dot a/a$ is the Hubble parameter. Introducing the slow-roll 
parameters,
\begin{eqnarray}
&&\epsilon={1\over2}\biggl({M_{\rm pl}V'\over V}\biggr)^2,
\\
&&\eta={M^2_{\rm pl}V''\over V},
\end{eqnarray}
we may write the the spectral index and tensor-to-scalar ratio as
\begin{eqnarray}
  &&n_s=1-6\epsilon+2\eta,
  \label{defns}
\\
&&r=16\epsilon,
\end{eqnarray}
and the energy density during the inflation epoch is written as
$\rho=(1+\epsilon/3)V$. 
We define the end of inflation as $\epsilon=1$, at which
the energy density of the universe can be written as 
\begin{eqnarray}
  \rho_{\rm end} ={4\over 3}V(\phi_{\rm end})={4\over 3}V_{\rm end},
\label{rhoend}
\end{eqnarray}
where $\phi_{\rm end}$ is the value of the scalar field at the end of inflation.
The $e$-folding number between horizon crossing of a perturbation of wavenumber
$k$ and the end of inflation is estimated as
\begin{eqnarray}
N_k=\ln\biggl({a_{\rm end}\over a_{k}}\biggr)
=-{1\over M_{\rm pl}^2}\int_{\phi}^{\phi_{\rm end}}{V\over V'}d\phi,
\label{defNk}
\end{eqnarray}
where $a_k$ and $a_{\rm end}$ are the scale factors at horizon crossing of 
a perturbation of wavenumber $k$ and at the end of inflation, respectively
(see figure \ref{figeras}).

Following previous works \cite{Dai,Munoz,Cook,Cai}, we first assume that  
during the reheating epoch, the universe is dominated by an energy component with an effective 
equation-of-state parameter $w_{\rm re}$. At the end of the reheating epoch, we assume that the energy 
density of the universe is written as
\begin{eqnarray}
  \rho_{\rm re}={\pi^2 g_{\rm re}\over 30} T_{\rm re}^4,
\label{rhore}
\end{eqnarray}
where $T_{\rm re}$ is the reheating temperature, and $g_{\rm re}$ is the
number of internal degrees of freedom of relativistic particles at the
end of reheating, which we assume to be $g_{\rm re}={\cal O}(100)$.
Defining the scale factor at the end of reheating, $a_{\rm re}$, then,
we can write the $e$-folding number of the reheating epoch,
\begin{eqnarray}
N_{\rm re}
&=&\ln\biggl({a_{\rm re}\over a_{\rm end}}\biggr)
=-{1\over 3(1+w_{\rm re})}
\ln\biggl(\frac{\rho_{\rm re}}{\rho_{\rm end}}\biggr),
\label{2.5.2}
\end{eqnarray}
where $a_{\rm re}$ is the scale factor at the end of reheating.

Using an observational constraint on the spectral index of the initial curvature 
perturbations, we can constrain the $e$-folding number $N_{\rm re}$ and 
the effective equation-of-state parameter $w_{\rm re}$ of the reheating epoch. 
The horizon crossing of a perturbation with the wavenumber $k$ occurs at 
$a_kH_k=k$, where $a_k$ and $H_k$ are the scale factor and Hubble parameter, respectively,
at horizon crossing during inflation. Then, we can write
\begin{eqnarray}
0=\ln\biggl({k\over a_kH_k}\biggr)=\ln\biggl({a_{\rm end}\over a_k}{a_{\rm re}\over a_{\rm end}}
{a_0\over a_{\rm re}}{k \over a_0 H_k}\biggr),
\label{NNN0}
\end{eqnarray}
where $a_0$ is the scale factor at the present epoch. Using the definitions Eqs.~(\ref{defNk})
and (\ref{2.5.2}), Eq.~(\ref{NNN0}) yields
\begin{eqnarray}
N_k+N_{\rm re}+\ln\biggl({a_{0}\over a_{\rm re}}\biggr)+\ln\biggl({k\over a_0H_k}\biggr)=0. 
\label{NNN2}
\end{eqnarray}
The geometrical meaning of Eq.~(\ref{NNN2})  is the equality in the lengths $\ell_x=\ell_y$
in the left panel of figure \ref{figeras}.

From the conservation of entropy, we may write
\begin{eqnarray}
{a_{\rm re}\over a_0}=\left(\frac{43}{11g_{re}}\right)^{{1}/{3}}{T_0\over T_{\rm re}},
\label{2.5.4}
\end{eqnarray}
where $T_0=2.725$~K is the temperature of the universe at the present epoch. Using 
Eq.~(\ref{rhore}), Eq.~(\ref{2.5.4}) is rewritten as 
\begin{eqnarray}
{a_{\rm re}\over a_0}=\left(\frac{43}{11g_{re}}\right)^{{1}/{3}}{T_0}
\left({\pi^2 g_{re}\over 30 \rho_{re}}\right)^{1/4}. 
\label{2.5.X}
\end{eqnarray}
Furthermore, using Eqs.~(\ref{rhoend}) and (\ref{2.5.2}), we have
\begin{eqnarray}
  \rho_{re}={4\over 3}V_{\rm end}\left({a_{\rm re}\over a_{\rm end}}\right)^{-3(1+w_{\rm re})}=
{4\over 3}V_{\rm end}e^{-N_{\rm re}3(1+w_{\rm re})}.
\end{eqnarray}
Then, the logarithm of Eq.~(\ref{2.5.X}) yields the following expression in terms of $N_{\rm re}$: 
\begin{eqnarray}
\ln\left({a_{\rm re}\over a_0}\right)={1\over 3}\ln\left(\frac{43}{11g_{re}}\right)
+{1\over 4}\ln\left({\pi^2 g_{re}\over 30}\right)
+{1\over 4}\ln\left({3T_0^4\over 4 V_{\rm end}}\right)+{3N_{\rm re}(1+w_{\rm re})\over 4}. 
\label{2.5.XX}
\end{eqnarray}

Using the amplitude of the scalar perturbations, $A_s=H^4/(4\pi^2\dot\phi^2)$,
and the slow-roll approximation, we may write
\begin{eqnarray}
H_{k}=\frac{\pi M_{\rm pl}\sqrt{r A_s}}{\sqrt{2}}.
\label{2.5.6}
\end{eqnarray}
Inserting Eqs.~(\ref{2.5.XX}) and (\ref{2.5.6}) into Eq.~(\ref{NNN2}), we finally have
\begin{eqnarray}
&&N_{\rm re}=\frac{4}{1-3w_{\rm re}}\Bigg[-N_{k}-\ln\left(\frac{k}{a_{0}T_{0}}\right)-\frac{1}{4}\ln\left(\frac
{40}{\pi^2g_{\rm re}}\right)-\frac{1}{3}\ln\left(\frac{11g_{re}}{43}\right)\nonumber\\
&&~~~~~~~~
+\frac{1}{2}\ln\left(\frac{\pi^2M_{\rm pl}^2\:r\:A_{s}}{2V_{\rm end}^{1/2}}\right)\Bigg]. 
\label{2.5.7}
\end{eqnarray}
In our analysis, we adopt 
the amplitude of the scalar perturbation at the pivot scale $A_s$ 
given by $10^{10}A_{s}=e^{3.064}$ (Table 4 of \cite{Planckparameter})
and $k=0.05~{\rm Mpc}^{-1}$ as a pivot wavenumber. 
Combining Eqs.~(\ref{rhoend}), (\ref{2.5.2}), and (\ref{2.5.4}), we also have
\begin{eqnarray}
&&T_{\rm re}=\mathrm{exp}\Bigg[-\frac{3}{4}(1+w_{\rm re})N_{\rm re}\Bigg]
\left(\frac{2V_{\rm end}}{5\pi^2}\right)^{1/4}.
\label{2.5.8}
\end{eqnarray}
Because the wavenumber $k$ and $n_s$ are related implicitly through the scalar field
$\phi$ with $H_ka_k=k$, Eqs.~(\ref{defns}) and (\ref{defNk}), 
one can write $N_{\rm re}$ and $T_{\rm re}$ as
functions of the spectral index $n_s$.

\section{Single-field $\alpha$-attractors}
In this paper, we focus on a class of single-field inflation models of the 
$\alpha$-attractors in a unified manner \cite{Galante,Scalisi,Linde2015}, 
which includes Starobinsky's $R^2$ inflation model \cite{Starobinsky,Mukhanov} 
and the Higgs inflation model \cite{Salopek,Bezrukov,Okada,Bezrukov2,LNW}.
This class of inflation models can be generated by 
spontaneously breaking the conformal symmetry \cite{SCI1,SCI3,Carrasco1,Carrasco2}.  
In this paper, we consider the E-model and T-model 
as generalised models of $\alpha$-attractors, which are
specified by the potential ~(\ref{Emodel}) and (\ref{Tmodel}), respectively.
Starobinsky's model corresponds to the E-model with $\alpha=1$ and $n=1$ in Eq.~
(\ref{Emodel}). 
Single power-law inflation models are reproduced as the limit of large $\alpha$. 

Ref.~\cite{Planck} demonstrates that the $\alpha$-attractor models are consistent 
with observations of the cosmic microwave background anisotropies. 
For consistency with the observed tensor-to-scalar ratio, roughly, we require that the 
parameter $\alpha$ is less than ${\cal O}(100)$. 
In figures \ref{EmodelNTmax}--\ref{fig:constraintonbeta}, the shaded region 
in each panel is excluded from the constraint on the scalar--tensor ratio.  
We first investigate constraints
on reheating after an inflation of the E-model and T-model by following  
previous works \cite{Dai,Munoz,Cook,Cai}. 
To this end, we adopt 
\begin{eqnarray}
n_s=0.9667\pm0.0040
\label{conns}
\end{eqnarray}
(see Table 4 in Ref.~\cite{Planckparameter}).

\subsection{E-model}
The E-model is specified by the potential \cite{Carrasco1,Carrasco2} 
written as
\begin{eqnarray}
V=\Lambda^4\biggl(1-e^{-\sqrt{\frac{2}{3\alpha}}{\frac{\phi}{M_{\rm pl}}}}\biggr)^{2n},
\label{Emodel}
\end{eqnarray}
where $\Lambda$, $n$, and $\alpha$ are the parameters.
Using the slow-roll approximation, we find the expressions 
for the spectral index and the tensor-to-scalar ratio, 
\begin{eqnarray}
&&n_s
=1-{8n\Bigl(e^{\sqrt{2\over 3\alpha}{\phi\over M_{\rm pl}}}+n\Bigr)
\over3\alpha \Bigl(e^{\sqrt{2\over 3\alpha}{\phi\over M_{\rm pl}}}-1\Bigr)^2 },
\\
&&r
={64n^2\over 3\alpha \Bigl(e^{\sqrt{2\over 3\alpha}{\phi\over M_{\rm pl}}}-1\Bigr)^2 },
\end{eqnarray}
and for the $e$-foldings as functions of $\phi$ from Eq.~(\ref{defNk}),
\begin{eqnarray}
&&N_k
=-{3\alpha\over 4n}\biggl[
e^{\sqrt{2\over 3\alpha}{\phi_{\rm end}\over M_{\rm pl}}}-e^{\sqrt{2\over 3\alpha}{\phi\over M_{\rm pl}}}
+\sqrt{2\over 3\alpha}{\phi-\phi_{\rm end}\over M_{\rm pl}}
\biggr].
\end{eqnarray}
Thus, we can write $T_{\rm re}$ and $N_{\rm re}$ as functions of $n_s$, regarding 
$\phi$ as an implicit parameter.

\begin{figure}[t]
 \begin{center}
\hspace{0mm}\scalebox{1.00}{\includegraphics{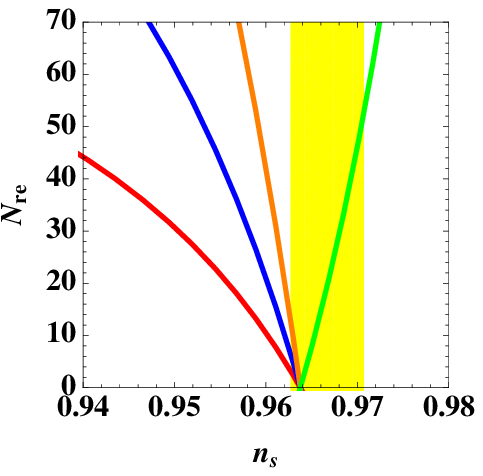}}
\hspace{0mm}\scalebox{1.00}{\includegraphics{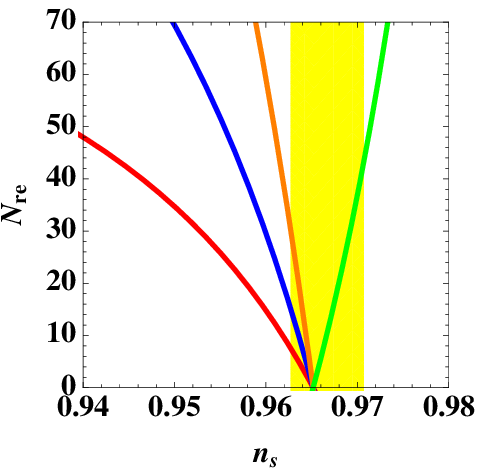}}
\hspace{0mm}\scalebox{1.00}{\includegraphics{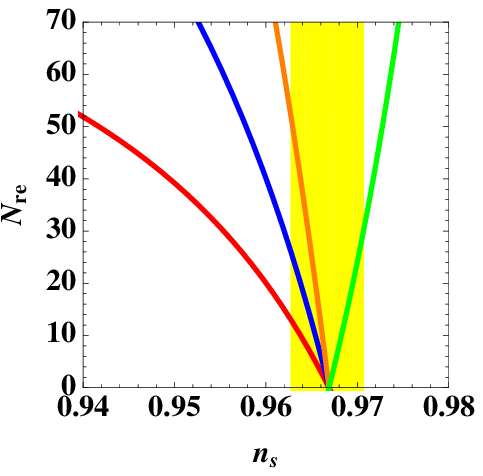}}
\\
\hspace{0mm}\scalebox{1.00}{\includegraphics{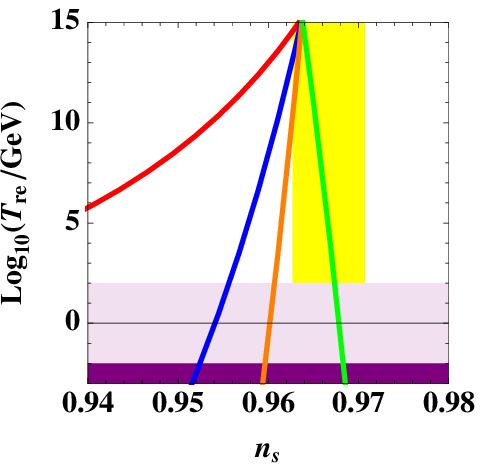}}
\hspace{0mm}\scalebox{1.00}{\includegraphics{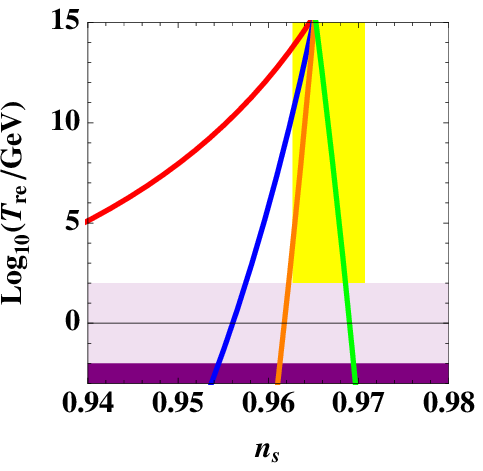}}
\hspace{0mm}\scalebox{1.00}{\includegraphics{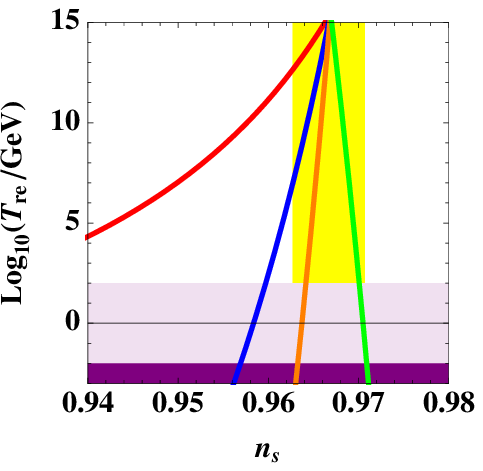}}
\caption{
$N_{\rm re}$ (upper panels) and $T_{\rm re}$  (lower panels) as functions of $n_s$ for 
E-model with $n=1$ and $\alpha=0.1$ (left panels), 
$\alpha=1$ (central panels), and $\alpha=5$ (right panels). 
In each panel, the curves represent different equation-of-state parameters
$w_{\rm re}$: $-1/3$ (red curve), 0 (blue curve), 1/6 (orange curve), and
2/3 (green curve). Yellow regions indicate the observational constraint, Eq.~(\ref{conns}).
In each panel, the point at which the four curves intersect, which corresponds to instant
reheating, gradually moves from left to right as the value of $\alpha$ increases.
Light purple and dark purple regions in lower panels show temperatures
below the electroweak scale, $T<\mathrm{100GeV}$, and the big bang nucleosynthesis 
scale, $T<\mathrm{10MeV}$, respectively. For consistency with big bang nucleosynthesis, 
$T_{\rm re}\gtrsim\mathrm{10 MeV}$.
\label{EmodelNT}}
\end{center}
\end{figure}

Figure \ref{EmodelNT} plots $N_{\rm re}$ (upper panels) and $T_{\rm re}$ (lower panels)
as functions of $n_s$, where we fix $n=1$. The left, central, and
right panels adopt $\alpha=0.1$, $1$, and $5$, respectively.
The curves in each panel represent different equation-of-state parameters $w_{\rm re}$:
$-1/3$ (red curve), 0 (blue curve), 1/6 (orange curve), and
2/3 (green curve). 
Our result for $n=1$ is the same as that in Ref.~\cite{Cook}. 

The yellow region shows the observational constraint on $n_s$, Eq.~(\ref{conns}).
Note that for a set of the parameter $w_{\rm re}$, $n$, and $\alpha$, there appears
the maximum value $N_{\rm re}^{(max)}$ so that the curve of $N_{re}$ is consistent
with the observational constraint on $n_s$. For example, in the central panels, which
assume $n=1$ and $\alpha=1$, the maximum value is $N_{\rm re}=8, ~15, ~30$,~and~$40$, respectively,
for $w_{\rm re}=-1/3, ~0, ~1/6$, and $2/3$.

\begin{figure}[t]
  \begin{center}
\hspace{0mm}\scalebox{1.00}{\includegraphics{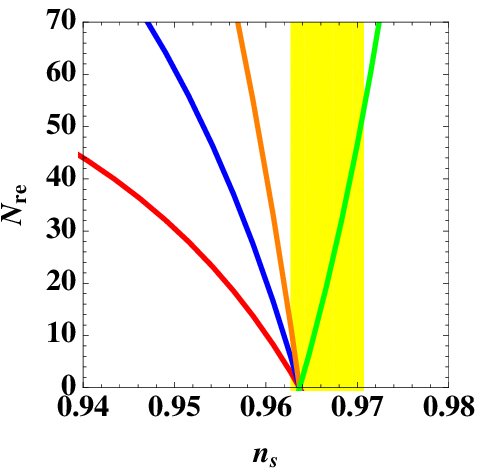}}
\hspace{0mm}\scalebox{1.00}{\includegraphics{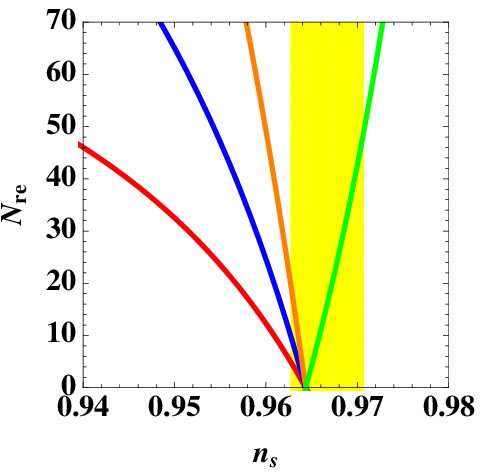}}
\hspace{0mm}\scalebox{1.00}{\includegraphics{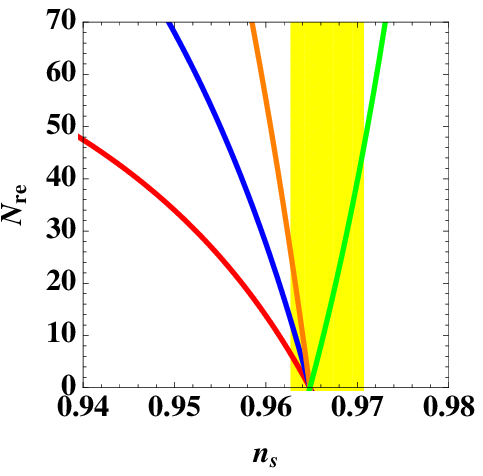}}
\\
\hspace{0mm}\scalebox{1.00}{\includegraphics{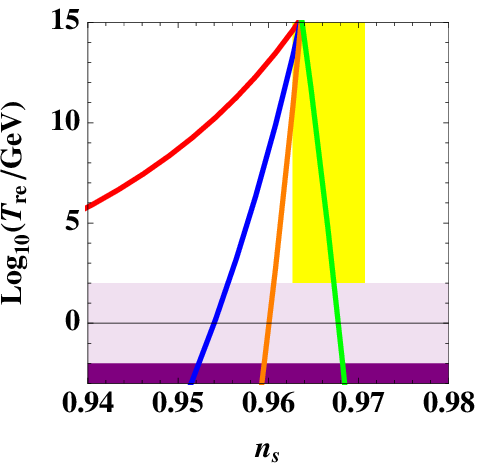}}
\hspace{0mm}\scalebox{1.00}{\includegraphics{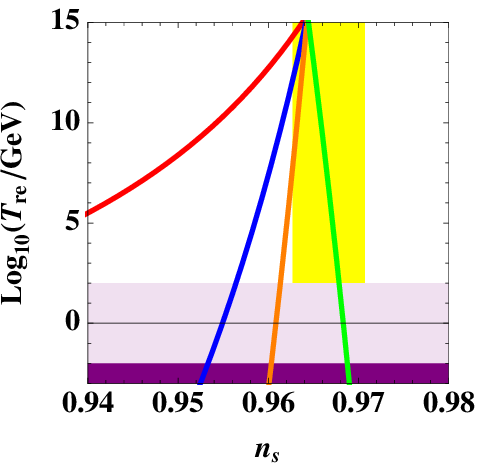}}
\hspace{0mm}\scalebox{1.00}{\includegraphics{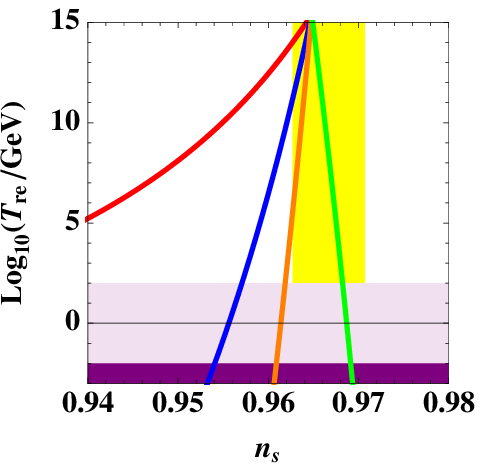}}
\caption{
  Same as figure \ref{EmodelNT} but for T-model. We fix $n=1$ and 
  $\alpha=0.1$ (left panels), $\alpha=1$ (central panels),
  and $\alpha=5$ (right panels).
  In each panel, the curves represent different equation-of-state parameters
$w_{\rm re}$: $-1/3$ (red curve), 0 (blue curve), 1/6 (orange curve), and
2/3 (green curve).
\label{TmodelNT}}
\end{center}
\end{figure}

\subsection{T-model}
The T-model is specified by the potential \cite{SCI3,Carrasco1,Carrasco2}
\begin{eqnarray}
V=\Lambda^4\tanh^{2n}\biggl({\phi\over \sqrt{6\alpha}{\rm M_{\rm pl}}}\biggr),
\label{Tmodel}
\end{eqnarray}
where $\Lambda$, $n$, and $\alpha$ are the parameters. Within the slow-roll 
approximation, we find expressions for the spectral index, 
tensor-to-scalar ratio, and $e$-foldings as functions of $\phi$:
\begin{eqnarray}
&&n_s=
1-{1\over 3\alpha}\biggl[8n(1+n){\rm csch}^2{\sqrt{2\over 3\alpha}{\phi\over M_{\rm pl}}}
+4n{\rm sech}^2{\sqrt{1\over 6\alpha}{\phi\over M_{\rm pl}}}\biggr],
\\
&&r=
{64n^2{\rm csch}^2{\sqrt{2\over 3\alpha}{\phi\over M_{\rm pl}}} \over 3\alpha },
\\
&&N_k=
-{3\alpha\over 4n}\biggl[
\cosh{\sqrt{2\over 3\alpha}{\phi_{\rm end}\over M_{\rm pl}}}
-\cosh{\sqrt{2\over 3\alpha}{\phi\over M_{\rm pl}}}\biggr].
\end{eqnarray}

Figure \ref{TmodelNT} is the same as figure \ref{EmodelNT} but for the T-model;
$N_{\rm re}$ (upper panels) and $T_{\rm re}$ (lower panels) are plotted
as functions of $n_s$, where we fix $n=1$. In the left, central, and
right panels, $\alpha=0.1$, $1$, and $5$, respectively.
In each panel, the curves represent different equation-of-state parameters
$w_{\rm re}$: $-1/3$ (red curve), 0 (blue curve), 1/6 (orange curve), and 
2/3 (green curve).

\begin{figure}[t]
\centering
\hspace{0mm}\scalebox{1.00}{\includegraphics{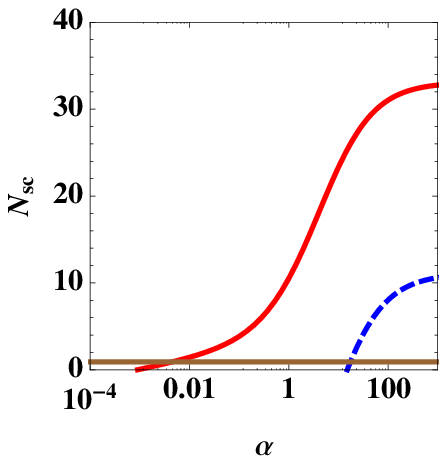}}
\hspace{0mm}\scalebox{1.00}{\includegraphics{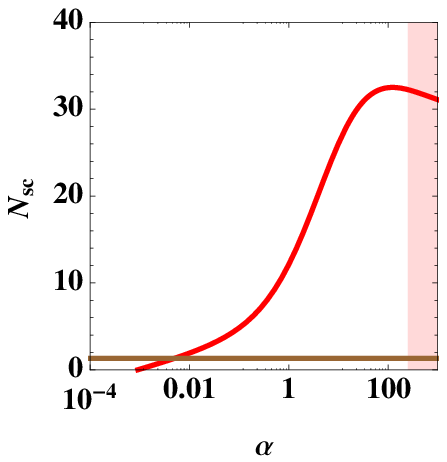}}
\hspace{0mm}\scalebox{1.00}{\includegraphics{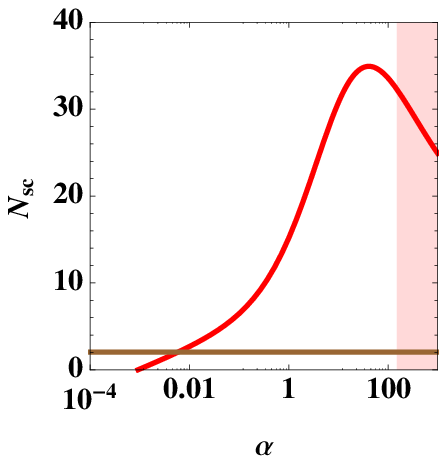}}
\\
\hspace{0mm}\scalebox{1.00}{\includegraphics{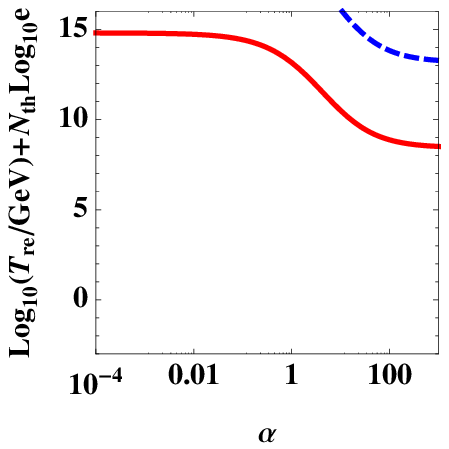}}
\hspace{0mm}\scalebox{1.00}{\includegraphics{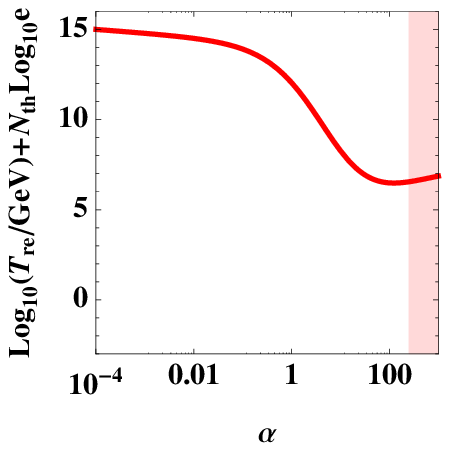}}
\hspace{0mm}\scalebox{1.00}{\includegraphics{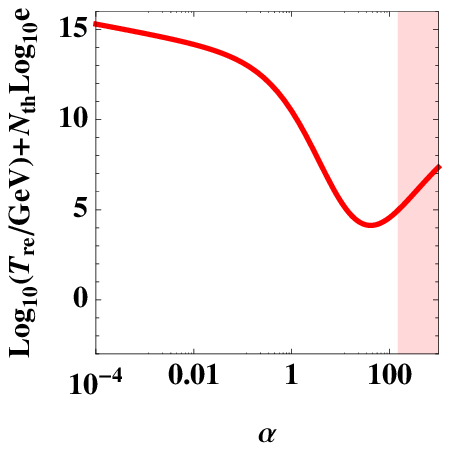}}
\caption{
The red solid curve is the maximum value of $N_{\rm sc}$ (upper panels) and the minimum value of
$\log_{10}(T_{\rm re}/{\rm GeV})+N_{\rm th}\log_{10}e$ (lower panels) as a function of $\alpha$ for
E-model with $n=1/2$ (left panel), $n=3/4$ (central panel), and $n=1$ (right panel).
Dashed curve in left panels shows the minimum value of $N_{\rm sc}$ 
and the maximum value of $\log_{10}(T_{\rm re}/{\rm GeV})+N_{\rm th}\log_{10}e$.
Brown line in upper panels shows the $e$-folding number for ${\cal N}_{\rm osc}=20$
oscillations, which is required for broad resonance preheating. 
Light shaded region in the central and right panels is excluded from 
the observational constraint on the tensor-to-scalar ratio $r$. 
\label{EmodelNTmax}}
\end{figure}

\begin{figure}[t]
\begin{center}
\hspace{0mm}\scalebox{1.00}{\includegraphics{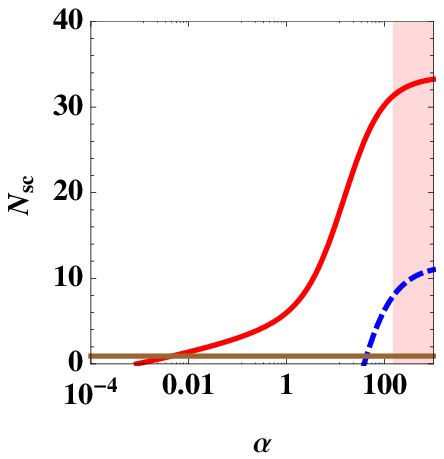}}
\hspace{0mm}\scalebox{1.00}{\includegraphics{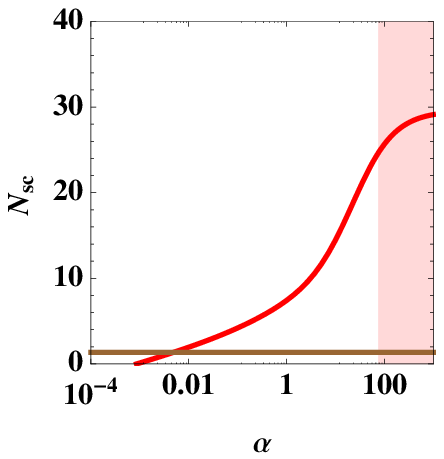}}
\hspace{0mm}\scalebox{1.00}{\includegraphics{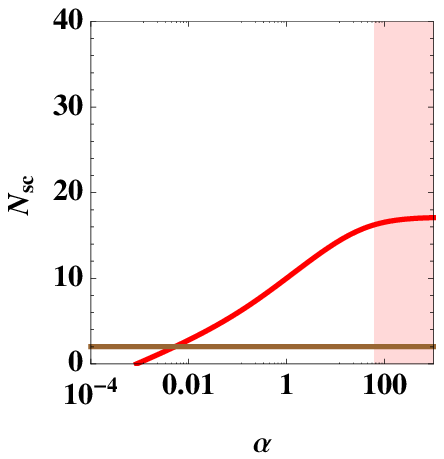}}
\\
\hspace{0mm}\scalebox{1.00}{\includegraphics{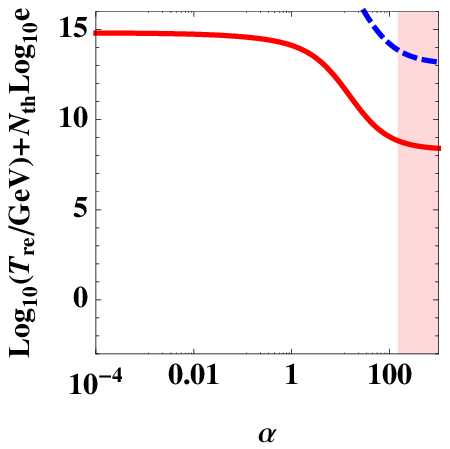}}
\hspace{0mm}\scalebox{1.00}{\includegraphics{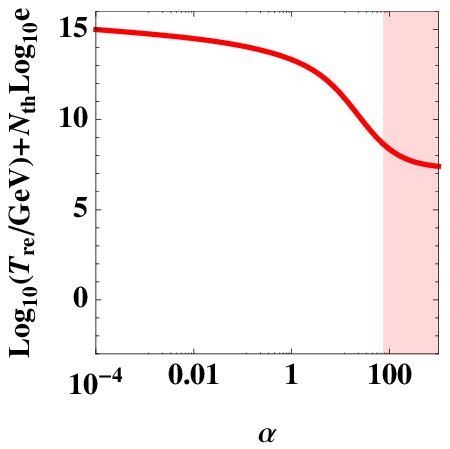}}
\hspace{0mm}\scalebox{1.00}{\includegraphics{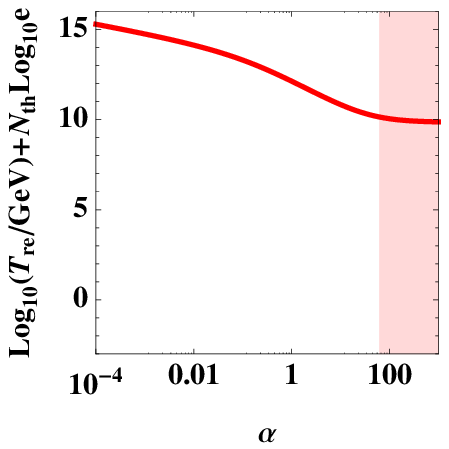}}
\caption{
Same as figure \ref{EmodelNTmax} but for T-model.
\label{TmodelNTmax}}
\end{center}
\end{figure}

\section{Two-phase reheating model}
In this section, we consider a simple scenario of reheating that
consists of two phases.
The first is an epoch in which the energy density of the universe is dominated by 
uniform oscillations of the inflaton field (the oscillation phase), and the second is 
an epoch in which the universe is dominated by relativistic but non-thermalised 
particles produced by decay of the inflaton field (the thermalisation phase). 
Figure \ref{figeras} illustrates the difference between the assumption of this 
section and that of the previous section. 
In the oscillation phase, the scalar field and scale factor follow 
Eqs.~(\ref{fieldequations2}) and (\ref{fieldequations1}), respectively.
When the scalar field oscillates around the minimum, which is approximated as
\begin{eqnarray}
V=\Lambda^4 \left({2\over 3\alpha M_{\rm pl}^2}\phi^2\right)^n
\label{VEmodel}
\end{eqnarray}
and
\begin{eqnarray}
V=\Lambda^4 \left({1\over 6\alpha M_{\rm pl}^2}\phi^2\right)^n
\label{VTmodel}
\end{eqnarray}
for the E-model and T-model, respectively, the equation-of-state parameter
of the scalar field is expressed in terms of the parameter $n$.
When the time scale of oscillation about the minimum is small, 
the virial theorem predicts that the energy density of the 
oscillating scalar field has the equation-of-state parameter 
\begin{eqnarray}
w_{\rm sc}={n-1\over n+1}
\label{wremm}
\end{eqnarray} 
for a potential $V\propto\phi^{2n}$ around the minimum \cite{Mukhanovtext}. 
One can check the validity of this formula by numerical solutions for
an expanding universe. This is because the period of oscillation is small 
compared to the Hubble time. 
Thus, the first oscillation phase of reheating is characterised by 
coherent oscillations of inflaton, in which the energy density is  
specified by the equation-of-state parameter $w_{\rm sc}$ in Eq.~(\ref{wremm}).

During the oscillation phase, light relativistic particles are 
produced gradually by a certain mechanism.
We assume that the energy density of the oscillating field and the energy
density of relativistic particles become equal at the scale factor
$a_{\rm eq}$ and that the relativistic particle component
dominates the energy density of the universe after $a_{\rm eq}$. 
However, the thermalisation process might not be completed quickly. 
Then, the second phase of
reheating is for thermalisation. We assume that the thermalisation 
phase continues until the scale factor becomes $a_{\rm re}$, at which the 
temperature of the universe is $T_{\rm re}$ and the energy density is
given by Eq.~(\ref{rhore}). Then, the $e$-folding number of the reheating epoch
is written as a combination of the two phases:
\begin{eqnarray}
N_{\rm re}
&=&\ln\biggl({a_{\rm re}\over a_{\rm end}}\biggr)
=\ln\biggl({a_{\rm re}\over a_{\rm eq}}{a_{\rm eq}\over a_{\rm end}}\biggr)
=N^{}_{\rm sc}+N^{}_{\rm th},
\label{NNN}
\end{eqnarray}
where we defined
\begin{eqnarray}
&&N_{\rm sc}
=\ln\biggl({a_{\rm eq}\over a_{\rm end}}\biggr)=-{1\over 3(1+w_{\rm sc})}
\ln\biggl(\frac{\rho_{\rm eq}}{\rho_{\rm end}}\biggr),
\\
&&N^{}_{\rm th}=\ln\biggl({a_{\rm re}\over a_{\rm eq}}\biggr)=-{1\over 4}
\ln\biggl(\frac{\rho_{\rm re}}{\rho_{\rm eq}}\biggr),
\end{eqnarray}
and $N^{}_{\rm sc}$ and $N^{}_{\rm th}$ are the $e$-folding numbers for the
oscillation phase and thermalisation phase, respectively.

On the basis of this assumption, we repeat the computation in section 2, 
which yields the following expressions instead of Eqs.~(\ref{2.5.7})
and (\ref{2.5.8}):
\begin{eqnarray}
&&N_{\rm sc}=\frac{4}{1-3w_{\rm sc}}\Bigg[-N_{k}-\ln\left(\frac{k}{a_{0}T_{0}}\right)-\frac{1}{4}\ln\left(\frac
{40}{\pi^2g_{\rm re}}\right)-\frac{1}{3}\ln\left(\frac{11g_{re}}{43}\right)\nonumber\\
&&~~~~~~~~
+\frac{1}{2}\ln\left(\frac{\pi^2M_{\rm pl}^2\:r\:A_{s}}{2V_{\rm end}^{1/2}}\right)\Bigg],
\label{2.5.7B}
\\
&&T_{\rm re}e^{N_{\rm th}}=\mathrm{exp}\Bigg[-\frac{3}{4}(1+w_{\rm sc})N_{\rm sc}\Bigg]
\left(\frac{2V_{\rm end}}{5\pi^2}\right)^{\frac{1}{4}}.
\label{2.5.8B}
\end{eqnarray}
Note that the expression for $N_{\rm sc}$ is equivalent to $N_{\rm re}$ in Eq.~(\ref{2.5.7})
and that the reheating temperature is modified by the $e$-folding number of thermalisation 
$N_{\rm th}$, but $T_{\rm re}e^{N_{\rm th}}$  is the same as the right-hand side of Eq.~(\ref{2.5.8}). 

As we described in the previous section, a maximum $e$-folding number appears 
for consistency with the constraint on $n_s$. 
From Eq.~(\ref{2.5.7B}), for the two-phase reheating model, we obtain the
maximum $e$-folding number for $N_{\rm sc}$, which is the same as that for $N_{\rm re}$,
when we fix $\alpha$, $n$, and $w_{\rm sc}$ instead of $w_{\rm re}$. 
In addition, from Eq.~(\ref{2.5.8B}), we obtain the minimum reheating temperature. Note that
$T_{\rm re}e^{N_{\rm th}}$ in the two-reheating-phase model is the same as the right-hand
side of Eq.~(\ref{2.5.8}); therefore, we obtain the minimum reheating temperature,
$\log_{10}T_{\rm re}+N_{\rm th}\log_{10}e$, in this case.

Figure \ref{EmodelNTmax} shows the maximum value of $N_{\rm sc}$ (upper panels) and
$\log_{10}(T_{\rm re}/{\rm GeV})+N_{\rm th}\log_{10}e$ (lower panels) as functions of $\alpha$
with $n=1/2$ (left panels), $n=3/4$ (central panels), and $n=1$ (right panels)
for the E-model. Figure \ref{TmodelNTmax} is the same as figure \ref{EmodelNTmax} but 
for the T-model. 
\section{Impact on reheating scenarios}
In this section, we discuss the impacts of the results in the previous section
on reheating scenarios by comparing the results with theoretical predictions. 
We may consider two types of interaction between an inflaton field $\phi$ and 
a light scalar field $\chi$, 
\begin{eqnarray}
L^{(4)}_I=-\frac{1}{2}\tilde{g}^2\phi^2\chi^2
\label{fourp}
\end{eqnarray}
and 
\begin{eqnarray}
L^{(3)}_I=-g\phi\chi^2,
\label{threep}
\end{eqnarray}
which describe $\chi$-particle production through the processes 
$\phi+\phi\rightarrow\chi+\chi$ and  $\phi\rightarrow\chi+\chi$, respectively, 
where $\tilde g$ and $g$ are their respective coupling constants.  

\subsection{Perturbative reheating}
We first consider a scenario in which effective resonant particle creation 
does not occur. We consider the perturbative reheating scenario as an 
elementary reheating scenario in which inflatons decay perturbatively through 
interaction (\ref{threep}). In this case, the evolution of the number 
density of inflatons is described by
\begin{eqnarray}
\frac{d(a^3n_{\phi})}{dt}=-\Gamma_{\phi\rightarrow\chi\chi}(a^3n_{\phi}),
\label{equationph}
\end{eqnarray}
where the decay rate $\Gamma_{\phi\rightarrow\chi\chi}$, described through
the interaction in (\ref{threep}), is 
\begin{eqnarray}
\Gamma_{\phi\rightarrow\chi\chi}={g^2\over 8\pi m_\phi},
\end{eqnarray}
where $m_\phi$ is the inflaton's mass.
Assuming that the background universe is dominated by the energy density 
of  inflaton oscillation, which might be treated as a fluid with 
the equation-of-state parameter $w_{\rm sc}$ in Eq.~(\ref{wremm}), the Friedmann equation is 
\begin{eqnarray}
{\dot a^2\over a^2}={\rho_{\rm end}\over 3M_{\rm pl}^2}\left({a\over a_{\rm end}}\right)^{-6n/(1+n)}.
\end{eqnarray}

\begin{figure}[t]
\begin{center}
\hspace{0mm}\scalebox{1.}{\includegraphics{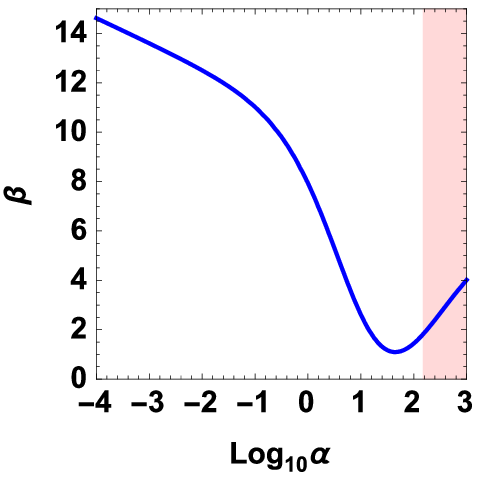}}
\hspace{15mm}\scalebox{1.}{\includegraphics{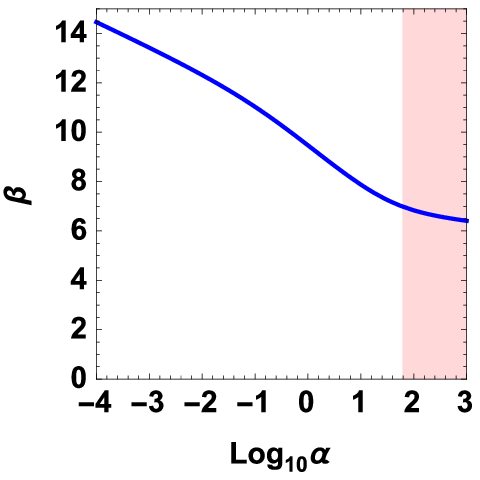}}
\caption{
The blue curve is the minimum value of $\beta$ for the coupling constant 
$g/\Lambda=10^{\beta-15}$~as a function of $\alpha$ for E-model 
(left panel) and T-model (right panel).
Here we adopted $n=1$ and $m_{\phi}=10^{13}$~GeV. 
Note that the coupling constant is defined so as to be $g=10^\beta$~GeV
when we choose $\Lambda=10^{15}$~GeV. 
\label{fig:constraintonbeta}}
\end{center}
\end{figure}
The above $e$-folding of perturbative reheating is simply understood 
as follows. 
We may estimate the epoch of $\chi$-particle decay as 
\begin{eqnarray}
H=\Gamma_{\phi\rightarrow\chi\chi},
\end{eqnarray}
which yields
\begin{eqnarray}
\left({a_{\rm eq}\over a_{\rm end}}\right)^{3n/(1+n)}
=\frac{8\pi m_{\phi}}{g^2M_{pl}}\sqrt{\frac{\rho_{\rm end}}{3}}
=\frac{16\pi m_{\phi}}{3g^2M_{pl}}{V_{\rm end}^{1/2}},
\end{eqnarray}
where we used Eq.~(\ref{rhoend}) in the second equality. 
We may write $V_{\rm end}\sim\Lambda^4\left({2}/{3\alpha}\right)^n$ 
and $V_{\rm end}\sim\Lambda^4\left({1}/{6\alpha}\right)^n$ for the E-model and T-model, respectively; 
then, we have the following expressions for the $e$-folding number, defined by $e^{N_{sc}}=a_{\rm eq}/a_{\rm end}$: 
\begin{eqnarray}
&&N_{sc}=-\frac{n+1}{3n}\ln{\bigg[\frac{3M_{pl}}{16\pi m_{\phi}}\left(\frac{g}{\Lambda}\right)^2
\left(\frac{3\alpha}{2}\right)^{n/2}\bigg]},
\label{Eqn.1}
\\
&&N_{sc}=-\frac{n+1}{3n}\ln{\bigg[\frac{3M_{pl}}{16\pi m_{\phi}}\left(\frac{g}{\Lambda}\right)^2
(6\alpha)^{n/2}\bigg]}
\label{Eqn.2}
\end{eqnarray}
for the E-model and T-model, respectively.
This puts a useful constraint on the coupling constant $g$ for a successful perturbative reheating 
scenario that is consistent with the observational constraint obtained in the previous section.
When we choose $m_{\phi}=10^{13}$~GeV 
defining 
\begin{eqnarray}
{g\over \Lambda}=10^{\beta-15},
\end{eqnarray} 
we have
\begin{eqnarray}
\beta\simgt 12.9-n\gamma-0.65{n\over n+1}N_{\rm sc}-{n\over 4}{\log_{10}\alpha}
\end{eqnarray}
where $\gamma=0.044$ and $\gamma=0.19$ for the E-model and T-model, respectively.
Figure \ref{fig:constraintonbeta}  shows the minimum value of $\beta$ as a function 
of $\alpha$ for 
the E-model (left panel) and T-model (right panel) with $n=1$.
For example, $\beta > 7.9$ for the E-model with $\alpha=1$ and $\Lambda=10^{15}$~GeV.
For the T-model with $\alpha=1$ and $\Lambda=10^{15}$~GeV, for successful 
perturbative reheating, $\beta>9.4$ is imposed. 
%

\subsection{Broad resonance preheating}
After the end of slow-roll inflation, the inflaton field $\phi$ oscillates
around a potential minimum, which is assumed to be approximated 
by Eqs.~(\ref{VEmodel}) and (\ref{VTmodel})
for the E-model and T-model, respectively. 
When $n=1$, these potentials are the harmonic potential, 
and we may assume that the oscillation of $\phi(t)$ is approximated by 
\begin{eqnarray}
\phi(t)\simeq\Phi\sin m_\phi t,
\end{eqnarray}
where $\Phi$ is the amplitude of the oscillation, and
$m_\phi$ is understood as $m_\phi=2\Lambda^2/\sqrt{3\alpha}M_{\rm pl}$ 
and $m_\phi=\Lambda^2/\sqrt{3\alpha}M_{\rm pl}$ and 
for the E-model and T-model, respectively. 
When $n\neq1$, the oscillation of $\phi(t)$ is not approximated 
by such a simple function. 

We here consider a resonant particle production scenario that 
was intensively investigated by Kofman, et al. \cite{Kofman} 
(see also \cite{Shtanov}). 
The equations of motion for a Fourier mode of the $\chi$ field are
\begin{eqnarray}
\ddot \chi_k(t)+3{\dot a\over a}\dot \chi_k(t)+\left({k^2\over a^2}
+m_\chi^2+\tilde g^2\phi^2(t)\right)\chi_k=0
\end{eqnarray}
and 
\begin{eqnarray}
\ddot \chi_k(t)+3{\dot a\over a}\dot \chi_k(t)+\left({k^2\over a^2}
+m_\chi^2+2g\phi(t)\right)\chi_k=0
\end{eqnarray}
for the interactions in Eqs.~(\ref{fourp}) and (\ref{threep}), respectively. 

This scenario of reheating relies on resonant particle creation due to the periodic time-dependent 
background at the earlier stage of reheating, which is called preheating. 
Particle creation effectively occurs when the Wentzel--Kramers--Brillouin approximation breaks down, which occurs at 
$\cos2m_\phi t\sim0$ or $\sin m_\phi t\sim 0$, depending on the interaction. 
We follow this scenario (see \cite{Mukhanov} for a review).
We first consider the four-point interaction. 
After ${\cal N}_{\rm osc}$ oscillations of inflaton field around the minimum,
the ratio of the number density of $\chi$ particles to that during inflaton is estimated as 
\begin{eqnarray}
\frac{n_{\chi}}{n_{\phi}}\sim\frac{k_{\ast}^3n_{k}({\cal N}_{\rm osc})}{\frac{1}{2}m_{\phi}\Phi_{0}^2}\sim m_{\phi}^{1/2}
\tilde{g}^{3/2}3^{{\cal N}_{\rm osc}}\Phi_{0}^{-1/2},
\end{eqnarray}
where we choose $k_*=m_\phi^3(\tilde g\Phi_0/m_\phi)^{3/2}$, and $\Phi_0$ is the inflaton's oscillation amplitude,
which we take to be $\Phi_0\sim M_{\rm pl}$. Using this relation, we can estimate the 
 ratio of the energy density of $\chi$  particles to that during inflation as
\begin{eqnarray}
\frac{\epsilon_{\chi}}{\epsilon_{\phi}}\sim\frac{m_{\chi}n_{\chi}}{m_{\phi}n_{\phi}}\sim\tilde{g}^{5/2} m_{\phi}^{-1/2}{\cal N}_{\rm osc}^{-1}3^{{\cal N}_{\rm osc}}\Phi_{0}^{1/2},
\end{eqnarray}
where we assumed $m_{\chi}={\cal O}(m_{\phi})={\cal O}(\tilde{g}\Phi)\sim{\tilde{g}\Phi_{0}}/
{{\cal N}_{\rm osc}}$. 
Then ${\epsilon_{\chi}}\simeq{\epsilon_{\phi}}$ appears after ${\cal N}_{\rm osc}$
oscillations, 
\begin{eqnarray}
{\cal N}_{\rm osc}\simeq 12\sim 30,
\end{eqnarray}
for a wide range of $10^{-5}<\tilde g<10^{-3}$, where we assumed $m_\phi= 10^{13}$~GeV. 
Here we compute the $e$-foldings to realise ${\cal N}_{\rm osc}=20$ inflaton oscillations; 
this yields the minimum duration required for successful preheating. 
The $e$-folding number for ${\cal N}_{\rm osc}=20$ oscillations is of order 
${\cal O}(1\sim2)$. Then we may write $N_{\rm sc}\simgt{\cal O}(1\sim 2)$. 
A more explicit value is obtained by solving Eqs.~(\ref{fieldequations2}) and (\ref{fieldequations1}).
$\chi$-particles can decay into other lighter particles quickly, which do not directly coupled to 
the inflaton \cite{Drewes}. 
The brown line in the upper panels of figures \ref{EmodelNTmax} and \ref{TmodelNTmax}
shows the value of $N_{\rm sc}$ for a broad resonance preheating scenario with 
${\cal N}_{\rm osc}=20$.
For the interaction in Eq.~(\ref{threep}), the estimation is essentially the same as 
the above estimation of the interaction in Eq.~(\ref{fourp}). 

When we also consider the constraint in the previous section, for consistency 
with the broad resonance preheating scenario, we need the rough condition
\begin{eqnarray}
\alpha\simgt 0.01. 
\end{eqnarray}

\section{Summary and Conclusions}
We investigated a constraint on the reheating epoch using an 
observational constraint on the spectral index $n_s$, in which  
we assumed the E-model and T-model as 
generalised $\alpha$-attractor models of inflation. 
When the reheating epoch is dominated by an energy component 
of the cosmic equation-of-state parameter $w_{\rm re}$, the $e$-folding 
number for reheating, $N_{\rm re}$, is bounded depending on 
$w_{\rm re}$, which also limits the reheating temperature $T_{\rm re}$.
Assuming that the reheating consists of two phases, an oscillation phase and 
a thermalisation phase, we investigated the $e$-folding number of the
oscillation phase $N_{\rm sc}$ and the reheating temperature 
$T_{\rm re}$,  depending on the equation-of-state parameter 
$w_{sc}$, which is determined by the potential. 
$N_{\rm sc}$ is constrained by the observational constraint on 
$n_s$, and the allowed regions of $N_{\rm sc}$ and $T_{\rm re}$ 
were obtained in section 4. 
For example, we found $N_{sc}\lesssim 16$ and 
$T_{re}e^{N_{\rm th}}\simgt 10^{10}{\rm GeV}$ for the E-model 
with $n=1$ and $\alpha=1$, whereas
$N_{sc}\lesssim 10$ and $T_{re}e^{N_{\rm th}}\simgt 10^{12}{\rm GeV}$
for the T-model with $n=1$ and $\alpha=1$. 
We discussed the implications of our results for two simple reheating scenarios. 
For the simplest perturbative reheating scenario, the ratio of the coupling constant $g$ 
for a decay to the mass scale of the potential of inflation $\Lambda$  should be 
$g\simgt10^{7.9}(\Lambda/10^{15}{\rm GeV})$~GeV for the E-model and 
$g\simgt10^{9.4}(\Lambda/10^{15}{\rm GeV})$~GeV for the T-model 
for $n=1$ and $\alpha=1$. 
Along a broad resonance preheating scenario, the $\alpha$ parameter is roughly constrained, 
$\alpha\simgt 0.01$, for the T-model and E-model.

\vspace{2mm}
\section*{Acknowledgment}
This work was supported by MEXT/JSPS KAKENHI Grant Number 15H05895.


\end{document}